# RED-2400: A Public Benchmark of Algorithmically-Rejected Trading Events with Outcome Labels

**Arati Uday Kamat**
Independent Researcher · ORCID: 0009-0000-4781-312X

## Abstract

RED-2400 is a public benchmark of 6,660 algorithmically-rejected trading events from a live Solana decentralised-exchange filter stack, observed continuously over 22 calendar days (2026-04-10T21:10Z through 2026-05-02T21:48Z, UTC). Of the 6,660 rejection events, 3,333 (50.05 percent) link exactly to a post-rejection price-and-liquidity sample series. The deposit contains 169,123 forward-outcome observations and 1,837 graveyard-tracker lifecycle snapshots, covering 1,076 distinct mints in the rejection registry and 1,075 in the forward-observation file. The deposit contains no outcome-label column. In its lifecycle-tracker file, `gone` is a tracker state recorded with liquidity 0; it is not terminal (167 of the 377 mints that reach it later return to an alive state) and does not denote token death or a rug pull. Filter labels are anonymised to `filter_1` through `filter_8`; source-collector identifiers to `source_a` and `source_b`. Liquidity and 24-hour volume are quantised to the nearest power of two, preserving heavy-tailed shape while preventing operational-threshold inference. This is the first window of a planned series; subsequent windows will extend the time horizon and enable regime-stratified analysis. "RED-2400" is a brand name, not a count; current cohort sizes are listed below and do not equal 2,400. Correction (October 2026): I withdraw Section V, which compared how often "early-death" and other rejected mints reach `gone`; 1,235 of the 1,236 early-death events were followed within 10.5 minutes by a re-rejection of the same token, so the label marks re-rejection timing, not token death. See the correction notice on page 1.

**Keywords:** filter design, counterfactual evaluation, reject inference, cryptocurrency, benchmark dataset, Solana, algorithmic decision systems.

**Correction notice (October 2026).** I withdraw Section V of the previous version of this paper and every conclusion drawn from it. Section V reported that 48.9 percent of "early-death" mints and 57.6 percent of other rejected mints reach the lifecycle state `gone`. The "early-death" label does not mark token death: 1,235 of the 1,236 early-death events were followed, within 10.5 minutes of their only sample, by a re-rejection of the same token, which restarted the tracker's record. The comparison therefore measures how often tokens are re-rejected, not whether they die. The Section V text is removed below, the sentences in Sections VI to VIII that depended on it are deleted, and items 3 and 5 of the dataset's v4 correction note, which rest on the same comparison, are withdrawn with it. Four further descriptions are corrected. (1) The post-rejection tracker sampled each tracked token every 10 minutes (99.51 percent of per-token sample gaps lie between 9.5 and 10.5 minutes), held one record per token and restarted it at each re-rejection; single-sample events and an event's last sample are therefore tracker artefacts, not token death, disappearance or survival. (2) `ageMin` is minutes since the rejection (0 to 1,440), not token age; 1,440 minutes is the tracker's horizon, not a cutoff for any outcome classification (Figure 2); and the row-level median of 686 minutes does not describe a typical event, whose last sample comes a median of 13 minutes after its rejection. (3) `gone` is a tracker state recorded with liquidity 0, not a sign that the price oracle stopped responding; it is not terminal (167 of the 377 mints that reach it later return to an alive state) and does not denote rug pull, fraud, token death or disappearance. `graveyard_lifecycle.csv` is therefore not lifecycle ground truth, and the use described in Section I, validating outcome labels against it, is not supported. (4) Only 3,333 of the 6,660 rejection events (50.05 percent) link exactly to a sample series; the statement that each event is linked is withdrawn. The Conflict of Interest statement is also corrected. The three data files are unchanged.

## I. Introduction

Filter-gated algorithmic trading systems on decentralised exchanges reject the majority of candidate tokens they evaluate. The post-rejection outcomes of those rejected candidates, the price and liquidity trajectories the system never participated in, are the empirical evidence base from which filter precision can be measured. RED-2400 is a public deposit of one such outcome record, observed continuously over a 22-day window on a live Solana trading deployment.

The deposit comprises three files. `rejections.csv` records the rejection events themselves with anonymised filter labels. `rejection_outcomes.csv` records the forward price-and-liquidity samples collected by a separate post-rejection tracking subsystem over a follow-up horizon of up to 24 hours per event. `graveyard_lifecycle.csv` records the lifecycle-state observations of rejected tokens collected by a third subsystem that polls each token for liveness, dormancy, and disappearance from the price oracle.

The principal evidentiary value of the deposit is the linkage between the rejection event and the lifecycle observation. With this linkage, researchers can test any outcome-classification rule, including the five-tier rule proposed by the methodology paper [Kamat 2026c], against the observed-lifecycle ground truth recorded in `graveyard_lifecycle.csv`. The deposit format is intended to support external validation as a primary use case.

## II. Data Description

### II.A File Listing

| File | Rows | Description |
|---|---|---|
| `rejections.csv` | **6,660** | One row per rejected decision event |
| `rejection_outcomes.csv` | **169,123** | Per-token forward time series linked to rejections |
| `graveyard_lifecycle.csv` | **1,837** | Lifecycle-state observations of rejected tokens |
| `checksums.txt` | n/a | SHA-256 manifest for the three CSV files |
| `LICENSE` | n/a | CC-BY-4.0 |
| `README.md` | n/a | Deposit description |

### II.B Observation Window

The deposit covers 2026-04-10T21:10:13Z through 2026-05-02T21:48:50Z, a span of **22 calendar days**. The graveyard-tracker subsystem began emitting on 2026-04-23T18:07Z and continued through 2026-05-02T16:53Z, providing approximately 9 days of lifecycle coverage for tokens already in the rejection registry.

### II.C Filter Distribution

Filter labels are anonymised to `filter_1` through `filter_8` in descending order of rejection volume:

| Filter | N | Share |
|---|---|---|
| filter_1 | 2,257 | 33.9 % |
| filter_2 | 1,530 | 23.0 % |
| filter_3 | 893 | 13.4 % |
| filter_4 | 843 | 12.6 % |
| filter_5 | 699 | 10.5 % |
| filter_6 | 321 | 4.8 % |
| filter_7 | 98 | 1.5 % |
| filter_8 | 19 | 0.3 % |
| **Total** | **6,660** | 100 % |

(Top-5 filter share: 93.4 percent; bottom-3 filter share: 6.6 percent.)

### II.D Schema

`rejections.csv` columns: `timestamp`, `source`, `mint`, `symbol`, `reason`, `timeSlot`. Source identifiers are anonymised to `source_a` and `source_b`. `reason` takes values `filter_1` through `filter_8`. `timeSlot` is `normal` or `strong`.

`rejection_outcomes.csv` columns: `sampleTs`, `mint`, `symbol`, `rejectReason`, `rejectTs`, `ageMin`, `priceUsd`, `liquidity`, `volume24h`, `dexId`, `pairAddress`, `source`. Liquidity and `volume24h` are quantised to the nearest power of 2 (log2 binning).

`graveyard_lifecycle.csv` columns: `ts`, `mint`, `symbol`, `from`, `to`, `liquidity`, `ageDays`. The `to` column captures the new lifecycle state and contains exactly three distinct values in the deposit: `alive_active`, `alive_dormant`, `gone`. The `from` column captures the previous state and additionally includes the empty string for rows representing the first observed lifecycle state of a mint.

### II.E Summary Statistics (Table 1)

| Field | Median | Mean | p10 | p90 |
|---|---|---|---|---|
| `ageMin` | 686 | 692 | 104 | 1,287 |
| `priceUsd` | $6.5 \times 10^{-6}$ | $3.6 \times 10^{-3}$ | $1.9 \times 10^{-6}$ | $2.5 \times 10^{-4}$ |
| `liquidity` | 8,192 | 26,920 | 4,096 | 32,768 |
| `volume24h` | 262,144 | 939,700 | 32,768 | 2,097,152 |

## III. The "RED-2400" Name

"RED-2400" is a brand name retained from a pre-deposit announcement when the matched analytic subset was estimated at approximately 2,400 events. The name should be treated analogously to dataset identifiers such as MNIST or CIFAR-10, where the number reflects a historical reference rather than a current statistic. Current reproducible cohort sizes in this deposit are 6,660 raw events and 2,582 events in the 14-day analytical window used by the related methodology paper. Neither equals 2,400. Citers should use "RED-2400" as a dataset identifier and the specific cohort sizes above when reporting analytical claims.

## IV. Relationship to the Methodology Paper

A related paper [Kamat 2026c] introduces a five-tier classification rule for post-rejection outcomes and reports filter-precision results computed by applying that rule to its own companion deposit at Zenodo DOI 10.5281/zenodo.19987695. **That companion deposit and the present RED-2400 deposit are separate Zenodo records with different file content and different cohort definitions.** They should not be substituted for each other in replication work. To reproduce the methodology paper's reported numbers (2,402 events; conservative save-to-miss ratio of approximately 3.7:1 under the windowed-saves definition), use the methodology paper's own deposit and the audit script bundled with it. The present RED-2400 deposit covers a longer window, a larger event count, and additionally provides the lifecycle-tracker file that the methodology paper's deposit does not include.

## V. External Validation of the Five-Tier Rule (withdrawn)

> **Section V withdrawn (October 2026).** The analysis previously printed here, a matched comparison of how often "early-death" and other rejected mints reach `gone` (48.9 versus 57.6 percent), is withdrawn and removed, together with its conclusions about the 14.8:1 and 3.7:1 ratios of the methodology paper. The "early-death" label marks re-rejection timing, not token death. See the correction notice on page 1. The earlier text remains available as version 2 of arXiv:2605.12151.

## VI. Reproduction

Reproduction commands are bundled with the deposit README. Briefly, the row counts and the per-filter shares of `rejections.csv` reproduce in any pandas or Node.js environment that can read the CSV. The graveyard `to`-state vocabulary reproduces by inspection of `value_counts` on the `to` column.

## VII. Limitations

First, the observation window (22 days) is short relative to the timescales over which DEX market microstructure regimes shift. The 22-day window reflects continuous operation of a specific production deployment; longer windows are intended for subsequent dataset versions but are not within scope for this deposit.

Second, the deposit reflects a single trading-system deployment on a single chain (Solana) at a single public DEX surface. Cross-chain replication is among the principal extensions sought.

Third, the deposited `graveyard_lifecycle.csv` is dense only after 2026-04-23T18:07Z, so lifecycle observation for rejection events earlier than that date is partial.

## VIII. Conclusion

RED-2400 is a deposit of 6,660 algorithmically-rejected trading events with linked forward outcomes and lifecycle observations, observed over 22 calendar days on a live Solana decentralised-exchange trading system. The dataset name "RED-2400" is brand-only and does not equal any cohort size in the deposit.

---

## Figures

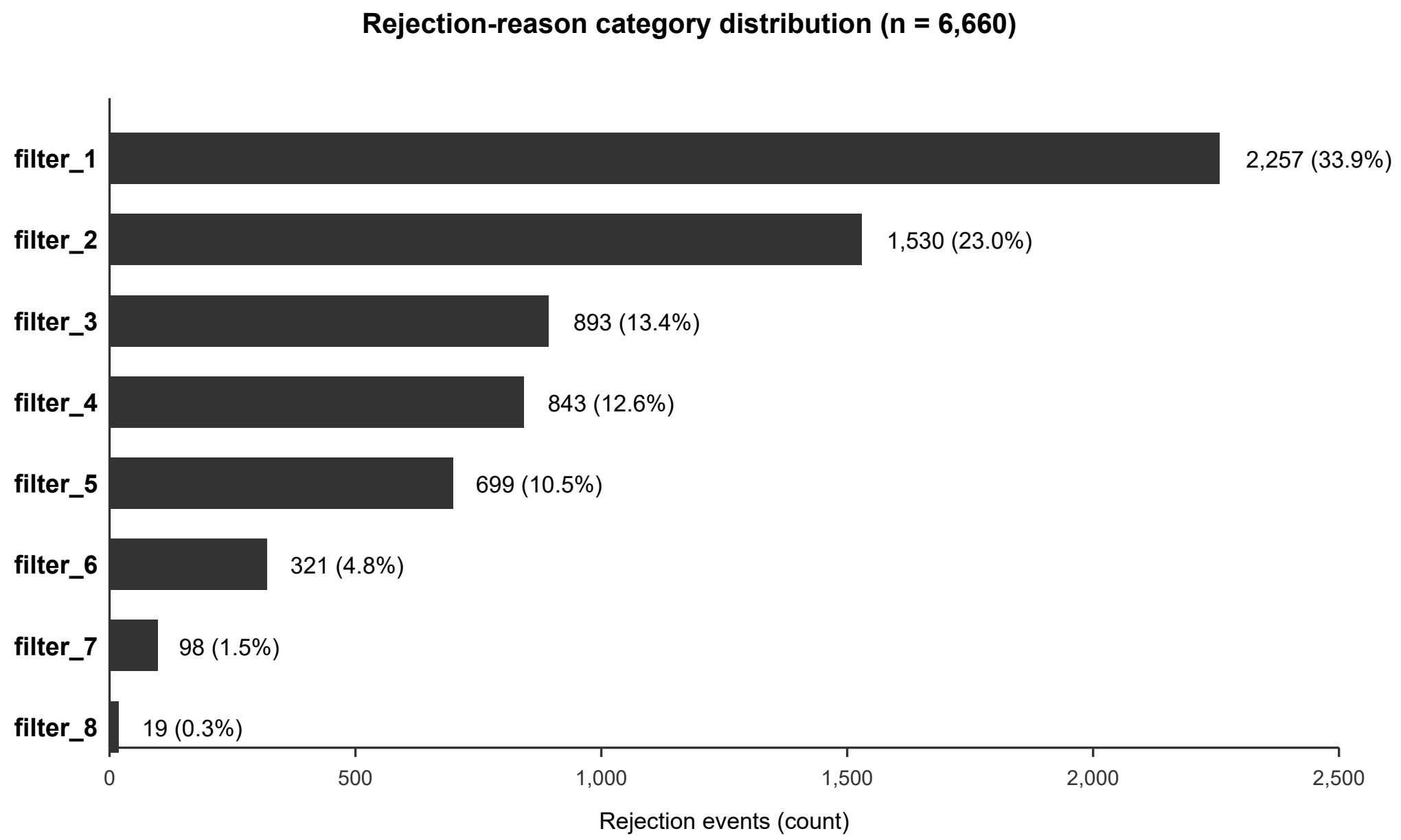

***Figure 1*** *Horizontal bar chart with raw counts and within-cohort percentages annotated to the right of each bar. The top five filters (filter_1 through filter_5) account for 93.4 percent of all rejections; filter_6 through filter_8 carry the remaining 6.6 percent. Volume rank is preserved by the anonymisation: filter_1 is the highest-volume rule and filter_8 the lowest.*

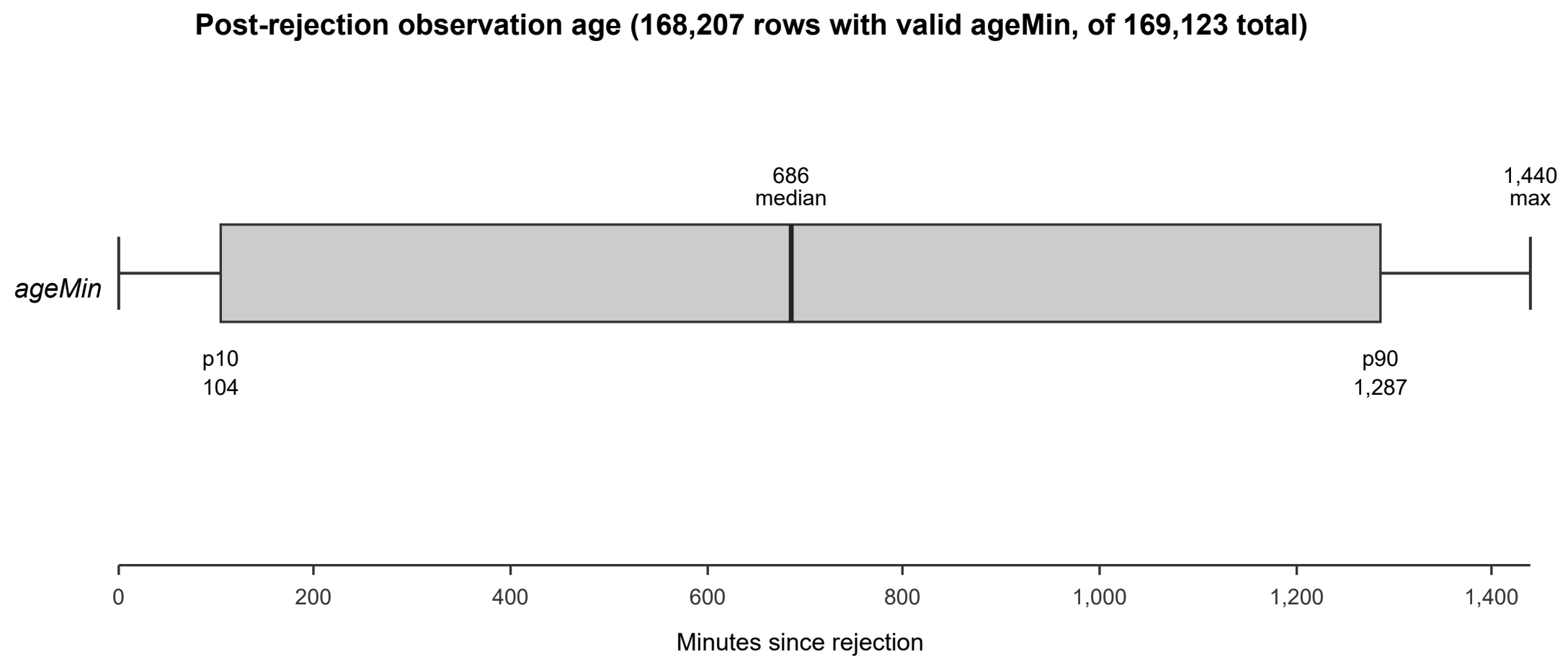

***Figure 2*** *Box plot of* `ageMin` *over the 168,207 rows with a valid (non-null, finite)* `ageMin` *value, out of 169,123 total forward-observation rows; the 916-row gap corresponds to outcome observations whose age field was null or non-numeric and is documented in* `PROVENANCE.md`*. The interquartile range is concentrated around the median of 686 minutes (approximately 11.4 hours), with p10 at 104 minutes and p90 at 1,287 minutes. The outcome time series for a typical rejected token covers roughly half of the 24-hour observation horizon. Right-censoring at 1,440 minutes (24 hours) marks the dataset cutoff for early-death classification.*

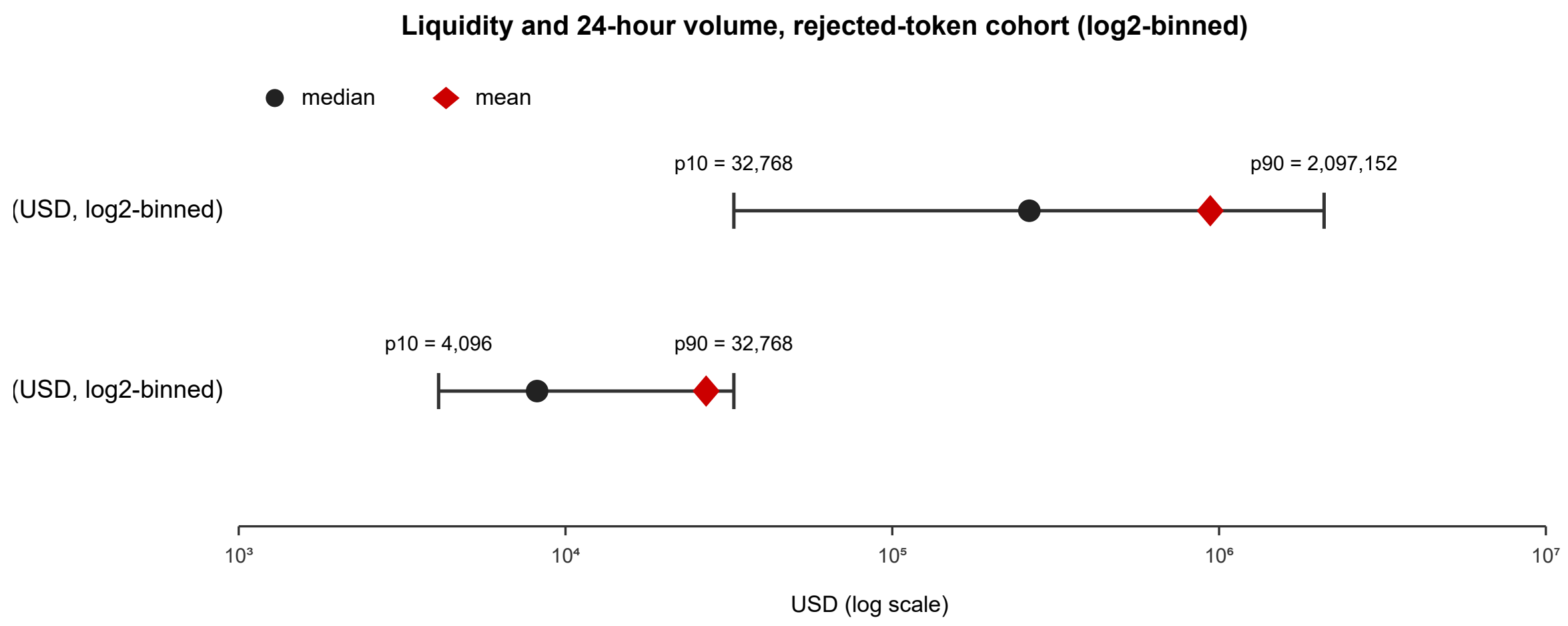

***Figure 3*** *Both fields use a log scale. The gap between median (black circle) and mean (red diamond) shows heavy-tailed structure on both axes. The bulk of rejected tokens trade with liquidity below USD 50,000; the filter stack is rejecting low-conviction venue activity. Values are reported on the log2-binned scale used in the public release.*

---

---

## Version History

- **Correction notice (October 2026):** Section V withdrawn and removed; dependent sentences in Sections VI to VIII deleted; abstract and competing-interests statement corrected. See page 1.
- **2.0 (2026-05-30):** Version corrected by the notice above. Companion dataset at Zenodo DOI 10.5281/zenodo.19989075.
- **1.0 (2026-05-02):** Superseded.

## Conflict of Interest

Corrected (October 2026): I developed the trading system whose rejections are recorded in this dataset and have pending U.S. provisional patent applications related to it.